\documentclass[aps,prl,secnumarabic,nobibnotes,twocolumn,superscriptaddress]{revtex4-1}
\usepackage{dsfont}
\usepackage{fontenc}
\usepackage{verbatim}
\usepackage{amsfonts}
\usepackage{mathrsfs}
\usepackage{amsmath}
\usepackage{color}
\usepackage{natbib}
\usepackage{graphicx}
\usepackage{bm}
\usepackage{amssymb}
\usepackage{xspace}
\usepackage{epstopdf}
\usepackage{dcolumn}
\usepackage{multirow}
\usepackage{booktabs}
\usepackage[colorlinks=true, letterpaper=true, pdfstartview=FitV, linkcolor=blue, citecolor=blue, urlcolor=blue]{hyperref}

\begin{document}

\title{Quadratic and Cubic Nodal Lines Stabilized by Crystalline Symmetry}

\author{Zhi-Ming Yu}
\email{Z.-M. Yu and W. Wu contributed equally to this work.}
\address{Research Laboratory for Quantum Materials, Singapore University of Technology and Design, Singapore 487372, Singapore}

\author{Weikang Wu}
\email{Z.-M. Yu and W. Wu contributed equally to this work.}
\address{Research Laboratory for Quantum Materials, Singapore University of Technology and Design, Singapore 487372, Singapore}

\author{Xian-Lei Sheng}
\address{Department of Physics, Key Laboratory of Micro-nano Measurement-Manipulation and Physics (Ministry of Education), Beihang University, Beijing 100191, China}
\address{Research Laboratory for Quantum Materials, Singapore University of Technology and Design, Singapore 487372, Singapore}

\author{Y. X. Zhao}
\email{zhaoyx@nju.edu.cn}
\address{National Laboratory of Solid State Microstructures and Department of Physics, Nanjing University, Nanjing 210093, China}
\address{Collaborative Innovation Center of Advanced Microstructures, Nanjing University, Nanjing 210093, China}

\author{Shengyuan A. Yang}
\email{shengyuan\_yang@sutd.edu.sg}
\address{Research Laboratory for Quantum Materials, Singapore University of Technology and Design, Singapore 487372, Singapore}

\begin{abstract}
In electronic band structures, nodal lines may arise when two (or more) bands contact and form a one-dimensional manifold of degeneracy in the Brillouin zone. Around a nodal line, the dispersion for the energy difference between the bands is typically linear in any plane transverse to the line. Here, we perform an exhaustive search over all 230 space groups for nodal lines with higher-order dispersions that can be stabilized by crystalline symmetry in solid state systems with spin-orbit coupling and time reversal symmetry. We find that besides conventional linear nodal lines, only lines with quadratic or cubic dispersions are possible, for which the allowed degeneracy cannot be larger than two. We derive effective Hamiltonians to characterize the novel low-energy fermionic excitations for the quadratic and cubic nodal lines, and explicitly construct minimal lattice models to further demonstrate their existence. Their signatures can manifest in a variety of physical properties such as the (joint) density of states, magneto-response, transport behavior, and topological surface states. Using \emph{ab-initio} calculations, we also identify possible material candidates that realize these exotic nodal lines.
\end{abstract}
\maketitle

Topological metals and semimetals have been attracting significant interest in current research~\cite{ChiuRMP2018,BansilRMP2016,ArmitageRMP2018}. These materials feature nontrivial degeneracies in their low-energy band structures, which give rise to novel fermionic excitations and lead to fascinating physical properties. According to the dimension of the degeneracy manifold, we may have zero-dimensional (0D) nodal points~\cite{WanPRB2011,MurakamiNJP2007,YoungPRL2012,YoungPRL2012,wangPRB2013}, 1D nodal lines (NLs), or even 2D nodal surfaces~\cite{Zhong2016NS,LiangPRB2016,WuPRB2018}. Of these possibilities, the NLs possess rich characteristics. For example, they exhibit a variety of topological connections in the Brillouin zone (BZ), such as isolated rings~\cite{YangPRL2014,MullenPRL2015}, loops traversing the BZ~\cite{ChenNL2015,LiNLPRB2017}, nodal chains~\cite{BzduvsekNat2016,WangNC2017,YuRuiPRL2017}, crossed nodal rings~\cite{WengPRB2015,YuRuiPRL2015,KimPRL2015}, nodal nets~\cite{ShengJPCL2017}, and Hopf links~\cite{ChenPRB2017,YanPRB2017,ChangPRL2017,ChangPRB2017}. Depending on the character of the two contacting bands (electronlike or holelike), the NL may also be classified as type-I, type-II, or hybrid type~\cite{LiNLPRB2017,ZhangPRB2018}.

There is yet another important characterization to classify a NL, namely, by the \emph{order} of energy dispersion around the line. To illustrate this, consider a NL formed by the contact between two bands. Around an arbitrary point $\bm K$ on the line, the effective Hamiltonian takes the generic form of
\begin{equation}\label{genericH}
\mathcal{H}_\text{eff}(\bm K+\bm q)=w(\bm q)+f(\bm q)\sigma_+ + f^*(\bm q)\sigma_- +g(\bm q)\sigma_z,
\end{equation}
where $\sigma_\pm=\sigma_x\pm i\sigma_y$ with $\sigma_{x,y,z}$ the three Pauli matrices, $w$, $f$, and $g$ are functions of the small wave-vector $\bm q$. Here, we are only concerned with the dispersion in the plane transverse to the NL, hence $\bm q$ is restricted to this plane. Also the $w$ term represents an overall energy shift for both bands, which does not affect the classification. Hence, more precisely, the classification is based on the leading order $\bm q$-dependence of $f$ and $g$. Since at $\bm q=0$, $f=g=0$, the typical dependence would be linear, which most NLs discussed to date belong to. Although NLs with quadratic dispersion were noticed in a few cases~\cite{ChangSR2017,ZhuPRX2016,LiLHPRB2017}, their symmetry requirement is not clear. Generally, to have leading-order dispersion beyond the linear one, it is necessary to require additional symmetries, such that their intricate interplay can eliminate the linear term. Then the natural question to ask is: \emph{Is it possible to have higher-order nodal lines, and what symmetries protect them?}

{\renewcommand{\arraystretch}{1.35}
\begin{table*}[t]
\centering{}%
\begin{tabular*}{18cm}{@{\extracolsep{\fill}}ccccccc}
\toprule
\toprule
SG No. & Point Group & Order  & Path   & IRR & SBF & ${\cal H}_{\rm eff}$\tabularnewline
\hline
174  & $C_{3h}$ & Quadratic & ${\rm \Gamma}$-A  & $\left\{ \Gamma_{4},\ \Gamma_{5}\right\} $  & $\left\{ |\frac{1}{2},\frac{1}{2}\rangle,\ |\frac{1}{2},-\frac{1}{2}\rangle\right\} $   & $\alpha q_-^2\sigma_+ +\text{H.c.}$
\tabularnewline
187$-$190  & $D_{3h}$ & Quadratic & ${\rm \Gamma}$-A   & $\Gamma_{4}$ & $\left\{ |\frac{1}{2},\frac{1}{2}\rangle,\ |\frac{1}{2},-\frac{1}{2}\rangle\right\} $ & $\alpha q_-^2\sigma_+ +\text{H.c.}$
\tabularnewline
183$-$186  & $C_{6v}$ & Cubic & ${\rm \Gamma}$-A  & $\Gamma_{9}$ & $\left\{ |\frac{3}{2},\frac{3}{2}\rangle,\ |\frac{3}{2},-\frac{3}{2}\rangle\right\} $ & $i(\alpha q_-^3+\beta q_+^3)\sigma_+ +\text{H.c.}$
\tabularnewline
\bottomrule
\bottomrule
\end{tabular*}\caption{List of all possible higher-order NLs stabilized by crystalline symmetry. IRR and SBF respectively stand for the irreducible representation and the standard basis function~\cite{Bradley1972}, which correspond to the two contacting bands that form the NL. It should be noted that in $\mathcal{H}_\text{eff}$, there may also be an overall shift term $w(q)=w_0+w_1q^2$ which does not affect the classification, and is therefore omitted. \label{tab:SG} }
\end{table*}
}


In this work, we address the above fundamental question by showing that higher-order NLs, including the quadratic and cubic lines, can be stabilized by crystalline symmetries. This is done by performing an exhaustive search over all 230 space groups (SGs) for solid state systems with spin-orbit coupling (SOC) and time reversal symmetry ($\mathcal{T}$). Our key results, as summarized in Table~\ref{tab:SG}, are the following. (i) NLs with quadratic and cubic dispersions, hereafter referred to as quadratic nodal line (QNL) and cubic nodal line (CNL), are the only stable ones beyond the conventional linear NLs. (ii) The degeneracy for all the identified higher-order NLs is two. This means that the corresponding system must have inversion symmetry ($\mathcal{P}$) broken, otherwise, the double degeneracy enforced by $\mathcal{PT}$ symmetry for each band necessarily make the NL (at least) four-fold degenerate. (iii) All the higher-order NLs are located along the high-symmetry path through the $\Gamma$ point, namely, $\Gamma$-A path, which indicates that the corresponding point group symmetry for each SG is actually  sufficient to stabilize the NLs. These results are further confirmed and explicitly demonstrated by the constructed effective $k\cdot p$ models and lattice models. We further show that the higher-order NLs feature distinct topological charges (Berry phases or winding numbers), which strongly influence their spectral, transport, and magnetic response properties (see Table \ref{tab:Exp_sign}). We also propose realistic materials which host such exotic NLs.

\textit{\textcolor{blue}{Rationale and effective Hamiltonian.}} We first describe the working procedure that leads to the result in Table~\ref{tab:SG}. A higher-order NL must require multiple symmetries for its stabilization, so it has to reside on the high-symmetry paths in the BZ. For each SG, we scan all the high-symmetry paths, looking for possible band degeneracies, which can be inferred from the dimensions of the irreducible representations for the little group on each path~\cite{Bradley1972}. Here, since SOC is fully considered, we deal with the double-valued representations, where a $2\pi$-rotation produces a minus sign and $\mathcal{T}^2=-1$. Then, for each irreducible representation with dimension $>1$, we construct the most general symmetry-preserving $k\cdot p$ Hamiltonian expanded at a generic point on the path, from which the order of the NL can be directly read off. This procedure is applied to all the 230 SGs, and the result is presented in Table~\ref{tab:SG}.

As an illustration, let us consider SG~174, which hosts a QNL along the $\Gamma$-A path (chosen as the $k_z$ axis). A generic point $\bm K$ on this path is invariant under the little group which contains two generators: the three-fold rotation $C_{3z}$ and the combined operation $\mathcal{T}M_z$, where $M_z$ is the reflection with the $x$-$y$ mirror plane. Since $[C_{3z},\mathcal{H}(\bm K)]=0$, the band eigenstates at $\bm K$ can be simultaneously chosen as $C_{3z}$ eigenstates. Consider a basis state $\psi_1$ corresponding to the $\Gamma_4$ representation of the little group, which transforms like $|1/2,1/2\rangle$. It is an eigenstate of $C_{3z}$, but is mapped to an orthogonal state $\psi_2\sim |1/2,-1/2\rangle$ under $\mathcal{T}M_z$.
It follows that $\{\psi_1, \psi_2\}$ gives a protected two-fold degeneracy on $\Gamma$-A.

{\renewcommand{\arraystretch}{1.35}
\begin{table*}[t!]
\centering{}%
\begin{tabular*}{18cm}{@{\extracolsep{\fill}}cccccccc}
\toprule
\toprule
 & Berry phase  & Phase offset & DOS & LL energy $\varepsilon_n(B)$  & Zero-energy LL    & JDOS  & Surface  State
 \tabularnewline
\hline
Nodal Point & $-$  & $\pm 1/8$ & $\sim|E|^2$ & $\sim (nB)^{1/2}$  & $-$  &   $\sim|E|^2$  & Fermi arc
\tabularnewline
Linear NL & $\pi$  & $0$ & $\sim|E|$ & $\sim (nB)^{1/2}$  &  1   &   $\sim|E|$  & drumhead
\tabularnewline
QNL & $0$   & $1/2$ & $\sim |E|^{0}$ & $\sim nB$  & 2  & $\sim |E|^{0}$  &  $-$
\tabularnewline
CNL & $\pi$ & 0  & $\sim|E|^{-\frac{1}{3}}$ $(\sim |E|^{0})^{\star}$   & $\sim (nB)^{3/2}$ $(\sim nB)^{\star}$ & 3   &  $\sim|E|^{-\frac{1}{3}}$ & span BZ
\tabularnewline
\bottomrule
\bottomrule
\end{tabular*}

\noindent \raggedright{}${\star}$ For CNLs, when $w_1q^2$ in the overall shift term dominates, the DOS and LL energy scaling relations would cross over to be with $\sim |E|^{0}$ and $\sim B$ dependence, respectively.

\caption{Comparison of properties between different NLs and also nodal point.  \label{tab:Exp_sign}}
\end{table*}
}

To construct the $k\cdot p$ effective Hamiltonian around $\bm K$, we express the symmetry operators in the $\{\psi_1, \psi_2\}$ basis:
\begin{equation}\label{Sym-174}
C_{3z}=e^{i\frac{\sigma_z}{2}\frac{2\pi}{3}},
\qquad
\mathcal{T}M_z=-i\sigma_x \mathcal{K},
\end{equation}
where $\mathcal{K}$ is the complex conjugation operator, and $\sigma_i$ are the Pauli matrices with $i=x,y,z$. The Hamiltonian is required to be invariant under the symmetry transformations, namely,
\begin{eqnarray}\label{C3}
&C_{3z}\mathcal{H}_\text{eff}(R^{-1}_{3z}\bm q)C_{3z}^{-1}=\mathcal{H}_\text{eff}(\bm q),&\\
\label{TM}
&(\mathcal{T}M_z)\mathcal{H}_\text{eff}(-\bm q)(\mathcal{T}M_z)^{-1}=\mathcal{H}_\text{eff}(\bm q),&
\end{eqnarray}
where, in line with the discussion for Eq.~(\ref{genericH}), $\bm q$ is in plane ($q_z=0$), and $R_{3z}$ is the three-fold rotation acting on $\bm q$.
Examining the Taylor series expansion of $g(\bm q)$ in Eq.~\eqref{genericH}, we see that all the even-order terms, which are invariant under the inversion of $\bm{q}$,  are excluded by Eq.~\eqref{TM} because $\{\mathcal{T}M,\sigma_z\}=0$. The linear term in $g(\bm{q})$ is also eliminated by the constraint of Eq.~\eqref{C3}, because $\sigma_z$ is invariant under $C_{3z}$, but $R_{3z}$ rotates $\bm{q}$. For the Taylor series of $f(\mathbf{q})$, the zeroth term vanishes because $C_{3z}\sigma_{\pm}C_{3z}^{-1}=e^{\pm i2\pi/3}\sigma_{\pm}$. To preserve $C_3$ symmetry, the linear term in $f(\bm{q})$ must take the form of $cq_+\sigma_++c^*q_{-}\sigma_{-}$, where $q_{\pm}=q_x\pm iq_y$ and $c$ is a constant.  However, such linear term is odd under $\mathcal{T}M_z$, hence it must vanish according to Eq.~\eqref{TM}. Thus, the linear terms in both $f$ and $g$ must vanish, and the corresponding effective Hamiltonian, to the leading order, reads
\begin{equation}\label{Heff174}
\mathcal{H}^{174}_\text{eff}(\bm q)=\alpha q_-^2\sigma_+ +\text{H.c.},
\end{equation}
where $\alpha$ is a complex parameter that depends on $K_z$. (Here and also in Eq.~\eqref{HeffCNL}, we have omitted the overall shift term $w(\bm q)$, for it does not affect band crossing.) It is worth to recall that the Taylor series of $g(\bm{q})\sigma_z$ begins at the third order. Hence, the effective Hamiltonian has an emergent chiral symmetry $\sigma_z$: $\{\mathcal{H}_{\rm eff}^{174}(\bm{q}),\sigma_z\}=0$, which holds up to the second order of $\bm{q}$.
Thus, we conclude that the doubly-degenerate line along $\Gamma$-A at leading order is a QNL with an approximate chiral symmetry.

The analysis for other SGs proceeds in a similar way. Particularly, for SGs~187-190, an additional mirror symmetry is required along with $C_{3z}$ and $\mathcal{T}M_z$. As a result, the effective Hamiltonians take the same form as Eq.~\eqref{Heff174}, but here $\alpha$ becomes real for SGs~187 and 188 with $M_x$, and purely imaginary for SGs~189 and 190 with $M_y$.

For SGs~183-186, the little group is generated by $C_{6z}$ and $M_x$. Most interestingly, they host CNLs along $\Gamma$-A, described by
\begin{equation}\label{HeffCNL}
\mathcal{H}^{183\text{-}186}_\text{eff}(\bm q)=i(\alpha q_-^3+\beta q_+^3)\sigma_+ +\text{H.c.},
\end{equation}
where $\alpha$ and $\beta$ are real parameters that depend on $K_z$. From our search, we find that no higher-order NL beyond CNL is protected by symmetry, and the degeneracy of all the identified QNLs and CNLs are two-fold.

To further confirm the existence of the QNLs and CNLs, we explicitly construct minimal lattice models for
the nine SGs listed in Table~\ref{tab:SG} (see Supplemental Material (SM) \cite{SM}). In each case, the lattice model reproduces the higher-order NL on the $\Gamma$-A path, and the model recovers the corresponding effective Hamiltonians when expanded around the path.

\textit{\textcolor{blue}{Topological charge.}} Conventionally, a nodal line can be regarded as an infinitely thin ``solenoid'' with fixed ``magnetic flux'' in momentum space, such that an electron circling around it along a closed path $C$ picks up a Berry phase similar to the Aharonov-Bohm effect~\cite{Berry1984}:
\begin{equation}
\gamma_C=\oint_C \mathcal{A}(\bm q) \cdot d\bm q \mod 2\pi,
\end{equation}
where $\mathcal{A}$ is the Berry connection for the occupied bands. Symmetries such as $\mathcal{PT}$ and mirror reflection can quantize $\gamma_C$ to be an integral multiple of $\pi$. Since a gauge transformation may change $\gamma_C$ by $2\pi$, it represents a $\mathbb{Z}_2$ topological invariant, namely, that $\gamma_C$ is defined mod $2\pi$. For the NLs under consideration, the QNLs and CNLs carry fluxes of $2\pi$ and $3\pi$, respectively, hence $\gamma_C$ is trivial for the QNLs, and is nontrivial for CNLs (if $|\alpha|\ne |\beta|$).

On the other hand, as aforementioned, all cases in Table~\ref{tab:SG} have an emergent chiral symmetry $\sigma_z$ with $\{\mathcal{H}_{\rm eff}(\mathbf{q}),\sigma_z\}=0$, which is exact up to the leading order. We can utilize this chiral symmetry to formulate another topological invariant~\cite{Volovik2003,SchnyderPRB2008,ZhaoPRL2013},
\begin{equation}\label{chiral-invariant}
\mathcal{N}=\frac{1}{4\pi i}\oint_C\mathrm{Tr}\ \sigma_z \mathcal{H}_{\rm eff}^{-1}(\bm{q})\nabla_{\bm{q}}\mathcal{H}_{\rm eff}(\bm{q}) \cdot d\bm{q},
\end{equation}
which is integer valued. Actually, we have the relation, $\gamma_C=\pi N\mod 2\pi$, for the two topological invariants.

For the QNLs, $\mathcal{N}=2$ to the leading order. If we consider the correction due to higher-order terms, for SG~174, there are terms $a(q_+^3+q_-^3)\sigma_z$ and $ib(q_+^3-q_-^3)\sigma_z$ at third order, with $a$ and $b$ real parameters, which violate the chiral symmetry. Consequently, Eq.~\eqref{chiral-invariant} is no longer quantized, but modified as $\mathcal{N}=2-2\pi q_c^2(a^2+b^2)/|\alpha|^2$, where $q_c$ is the radius of the circle $C$. Hence, $\mathcal{N}$ is a well-defined topological invariant for low-energy physics, as long as $q\ll |\alpha|/\sqrt{a^2+b^2}$. For SGs~187-188 (SGs~189-190), only the term $a(q_+^3+q_-^3)\sigma_z$ [$ib(q_+^3-q_-^3)\sigma_z$] exists, and $\mathcal{N}$ is modified accordingly. For the CNLs in SGs~183-186, we have $\mathcal{N}=3$ ($-3$) if $|\alpha|>|\beta|$ ($|\beta|>|\alpha|$). The next order nonvanishing term is $ic(q_+^6-q_{-}^{6})\sigma_z$ with $c$ a real parameter. The correction of $\mathcal{N}$ is proportional to $q^6$. Therefore, the emergent chiral symmetry is even more accurate for CNLs.

\textit{\textcolor{blue}{Experimental signature.}} The order of dispersion affects a variety of physical properties, leading to distinct signatures for the higher-order NLs. Some of these are listed in Table \ref{tab:Exp_sign} and compared with linear NLs and nodal points.

Particularly, the Berry phase can directly manifest when electrons are forced to circling around the NL under a magnetic field~\cite{MikitikPRL1999}. Consider a $B$ field along the $z$ direction. In the semiclassical picture, the electron orbits in momentum space are quantized according to the Bohr-Sommerfeld condition,
\begin{equation}\label{BSC}
\ell_B^2 A_{C_n}=2\pi (n+\nu),
\end{equation}
where $\ell_B=\sqrt{1/eB}$ is the magnetic length, $A_{C_n}$ is the area enclosed by the semiclassical orbit $C_n$ in $k$ space, $n$ is an integer corresponding to the Landau level (LL) index, and the phase offset $\nu$ is related to the Berry phase by $\nu=1/2-\gamma_{C_n}/(2\pi)$. In experiment, this orbit quantization leads to quantum oscillations in a variety of physical properties~\cite{Shoenberg1984}. For example, the Shubnikov-de Haas oscillation in magnetoresistance follows the relation $\delta\rho_{xx}\sim \cos[2\pi(\frac{F}{B}-\nu)]$, where $F$ is the oscillation frequency. Then, the phase offset $\nu$ and hence $\gamma_{C_n}$ can be extracted from experimental results by the standard analysis of the LL fan diagram. According to our discussion, $\nu$ should be $0$, $1/2$, and $0$ for the contribution from the linear, quadratic, and cubic NL fermions, respectively.

The order of dispersion also affects the scaling of the LL energy. From Eq.~(\ref{BSC}), we see that for an $m$-th order NL, $A_{C_n}\sim k^2\sim E^{2/m}$ for energy $E$, such that the LL energy scales as $\varepsilon_n \sim (n B)^{m/2}$ for large $n$. We have verified this relation by a full quantum solution of the LLs using the effective models in Table \ref{tab:SG}. The unconventional scaling in the LL spectrum offers another distinguishing property for higher-order NLs, which can be detected, \emph{e.g.}, by the scanning tunneling spectroscopy measurement. In addition, the order of the NL also determines the number of zero-energy LLs, where the topological charge of Eq.~\eqref{chiral-invariant} plays an essential role as discussed in SM~\cite{SM}.

There is caveat for the case of CNLs, due to the overall shift term $w(\bm q)$. This term, when expanded to third order for SGs~183-186, takes the form of $w_0+w_1 q^2$. It does not affect the classification, but the $w_1 q^2$ term may alter the LL scaling in a range where it dominates over the cubic term, for which the scaling would undergo a crossover from $\sim(nB)^{3/2}$ to $\sim (nB)$. Similar discussion applies to the scaling in the density of states. Nevertheless, the joint density of states (JDOS), which is crucial for optical transitions, is not affected by the shift term.

\begin{figure}
\includegraphics[width=8.4cm]{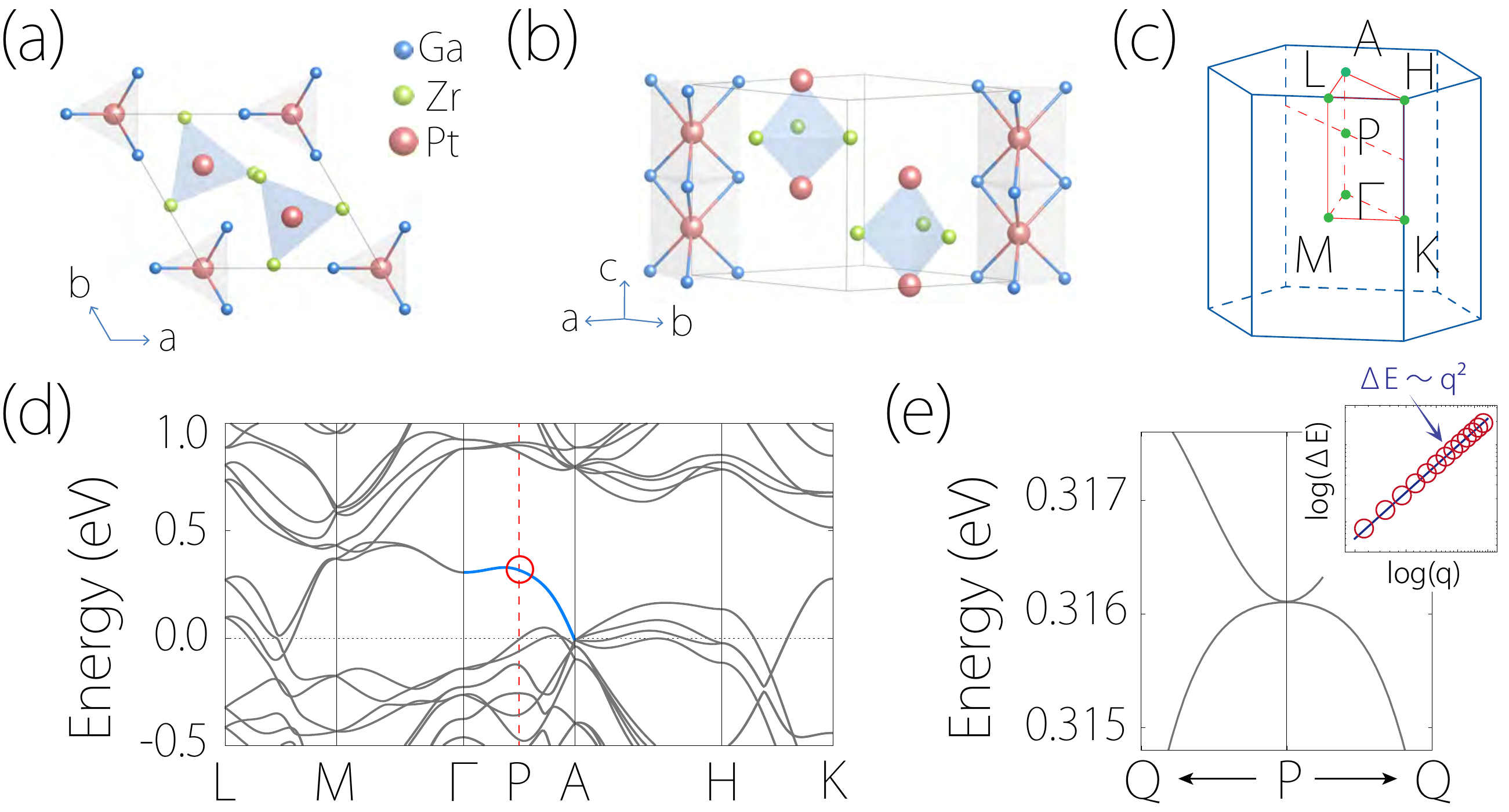}
\caption{(a) Top and (b) side view of the crystal structure for  ZrPtGa. (c) shows the BZ. (d) Calculated band structure for ZrPtGa with SOC. (e) Enlarged view of the band dispersion (in direction transverse to the NL) around a generic point P [marked in (d)]. P and Q are the mid-points of the paths $\Gamma$-A and K-H, respectively. The inset shows the log-log plot of the energy $\Delta E$ between the two bands versus $q$.
\label{Fig_material}}
\end{figure}

\textit{\textcolor{blue}{Material realization.}} To demonstrate that QNLs and CNLs can indeed exist in real materials, we have searched the existing materials with the target SGs and identified a few examples~\cite{SM}. One example is shown here. It is the intermetallic ternary compound ZrPtGa, which belongs to SG~190~\cite{Zumdick1999}. It has trigonal [Pt$_{2}$Ga$_{6}$] prisms running along the $c$ axis, with [Pt$_{2}$Zr$_{3}$] hexahedral clusters intercalated in-between [Fig.~\ref{Fig_material}(a,b)]. The calculated band structure for ZrPtGa (with SOC) is plotted in Fig.~\ref{Fig_material}(d). One finds that there is a NL along $\Gamma$-A with quadratic dispersion, more clearly observed in Fig.~\ref{Fig_material}(e) which shows a zoom-in image for the dispersion (perpendicular to the NL) around a generic point on the NL. The presence of this QNL is consistent with our symmetry analysis. The calculation details and more examples can be found in SM~\cite{SM}.

\begin{figure}[t!]
\includegraphics[width=8.1cm]{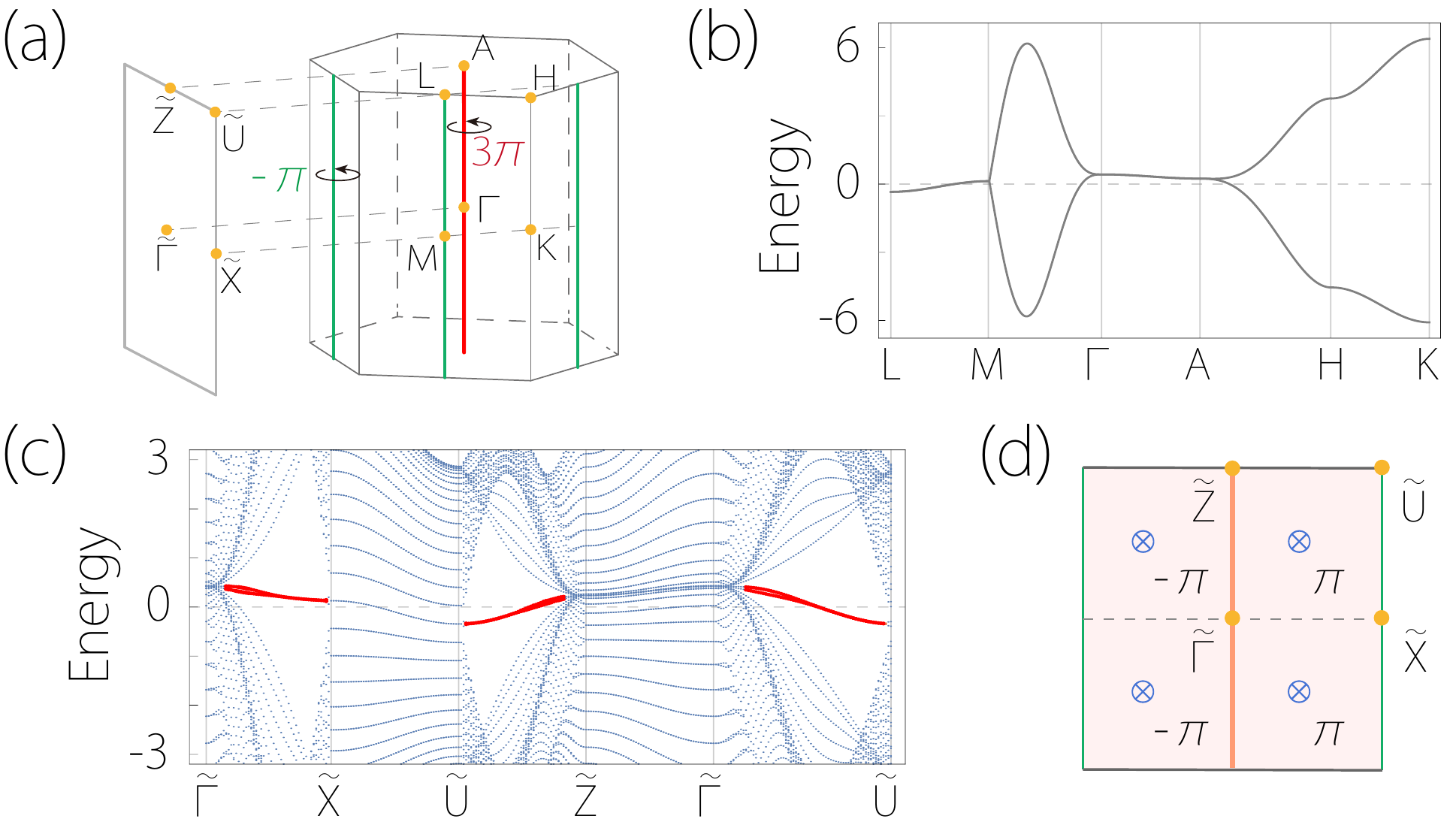}
\caption{(a) Bulk and surface BZs for SGs~183-186. (b) Bulk band structure for a lattice model with SG~183. Besides the CNL along $\Gamma$-A, there are three linear NLs along the three M-L paths, as indicated in (a). (c) shows the corresponding spectrum for a slab with (010) surfaces. The red color highlights the surface states, which span the whole surface BZ, as indicated in (d). $\pm \pi$ in (d) denote the Zak phase for lines perpendicular to the (010) surface. The calculation details are presented in SM~\cite{SM}.
\label{Fig_SS}}
\end{figure}

\textit{\textcolor{blue}{Discussion.}}
Some NL materials possess protected surface states. The protection is typically associated with a quantized $\pi$ Zak phase, which is the Berry phase defined for a straight line crossing the bulk BZ and perpendicular to the surface. The quantization requires additional symmetries, such as inversion symmetry or a reflection symmetry with the mirror plane parallel to the studied surface. For systems discussed here, we find that there is no protected surface state for the QNLs. Interestingly, for CNL systems, we find that surface states exist for the side surfaces, protected by nontrivial Zak phases with quantization enforced by the perpendicular mirror planes. Moreover, unlike the usual drumhead-like surface states~\cite{YangPRL2014,WengPRB2015}, these states span over the whole surface BZ, as illustrated in Fig.~\ref{Fig_SS}(c,d).

\begin{figure}[t]
\includegraphics[width=8.4cm]{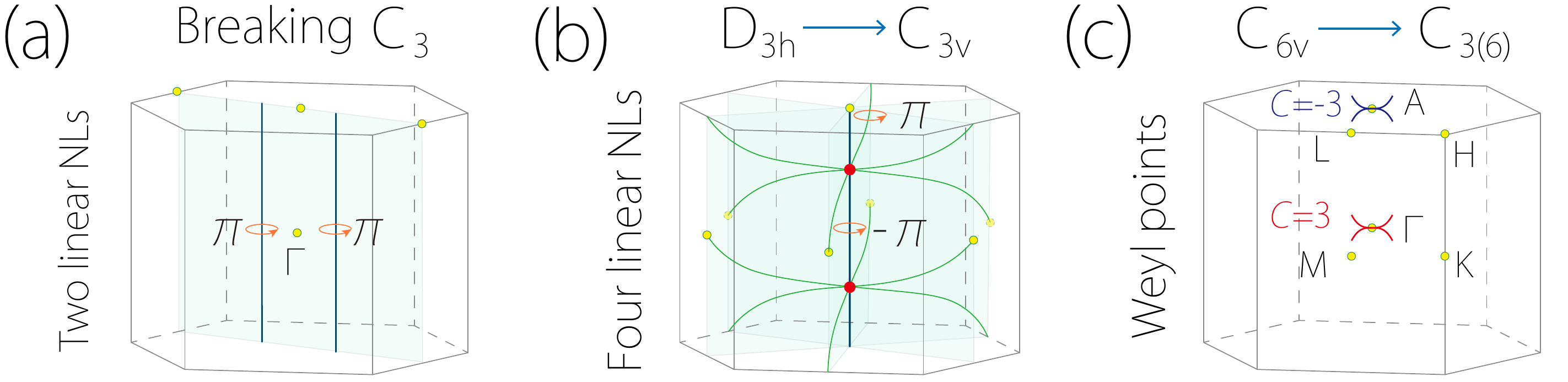}
\caption{QNL and CNL under symmetry breaking.  Breaking (a) $C_{3z}$ or (b) $M_z$  may transform the QNL of SG 187 into two linear NLs or four intertwined linear NLs (black and green curves), respectively. (c) Breaking $M_x$ may transform a CNL into two triple Weyl points at $\Gamma$ and A points  with opposite Chern number.  The calculation details are presented in SM~\cite{SM}.
\label{Fig_break}}
\end{figure}

Finally, as these NLs are protected by symmetry, they will transform under symmetry breaking. Some interesting cases are illustrated in Fig.~\ref{Fig_break}. For example, breaking $C_{3z}$ may split the QNL into two linear NLs [Fig.~\ref{Fig_break}(a)]; breaking $M_z$ for SG~174 will generally gap out the QNL, whereas for SGs~187-190, it may transform the QNL into four intertwined linear NLs [see Fig.~\ref{Fig_break}(b)]: One linear NL is along $\Gamma$-A, while the other three NLs penetrate through the BZ once along $z$ and twice along the in-plane directions. Moreover, breaking $M_x$ for SGs~183-186 will transform a CNL into two triple Weyl points [Fig.~\ref{Fig_break}(c)]. Thus, these systems provide a promising playground for studying topological phase transitions.

\bibliography{QCNL_ref}
\end{document}


\title{Supplemental Material for ``Quadratic and Cubic Nodal Lines Stabilized by Crystalline Symmetry''}

\author{Zhi-Ming Yu}
\address{Research Laboratory for Quantum Materials, Singapore University of Technology and Design, Singapore 487372, Singapore}

\author{Weikang Wu}
\address{Research Laboratory for Quantum Materials, Singapore University of Technology and Design, Singapore 487372, Singapore}

\author{Xian-Lei Sheng}
\address{Department of Physics, Key Laboratory of Micro-nano Measurement-Manipulation and Physics (Ministry of Education), Beihang University, Beijing 100191, China}
\address{Research Laboratory for Quantum Materials, Singapore University of Technology and Design, Singapore 487372, Singapore}

\author{Y. X. Zhao}
\address{National Laboratory of Solid State Microstructures and Department of Physics, Nanjing University, Nanjing 210093, China}
\address{Collaborative Innovation Center of Advanced Microstructures, Nanjing University, Nanjing 210093, China}

\author{Shengyuan A. Yang}
\address{Research Laboratory for Quantum Materials, Singapore University of Technology and Design, Singapore 487372, Singapore}



\maketitle
\maketitle
\section{Details of First-Principles Calculation}

We performed the first-principle calculations based on the density functional theory (DFT), as implemented in the Vienna \emph{ab initio} simulation package~\cite{Kresse1993,Kresse1996}.  The projector augmented wave method was adopted~\cite{Blochl1994}. The generalized gradient approximation (GGA) with the Perdew-Burke-Ernzerhof (PBE) realization~\cite{Perdew1996} was adopted for the exchange-correlation potential. For all calculations, the energy and force convergence criteria were set to be $10^{-6}$~eV and $10^{-2}$~eV/\text{\AA}, respectively. The BZ sampling was performed by using $k$ grids with a spacing of $2\pi \times 0.02$ \AA$^{-1}$ within a $\Gamma$-centered sampling scheme. For V$_{12}$P$_{7}$, the optimized structure paramemters ($a = 9.30$ \text{\AA}, $c = 3.22$ \text{\AA}) were used for the band structure calculation. For the other materials discussed below, the experimental values of their respective lattice parameters were used in the calculation. \\

\section{Material candidates}\label{cubic}

\begin{figure*}
\centering{}\includegraphics[width=8cm]{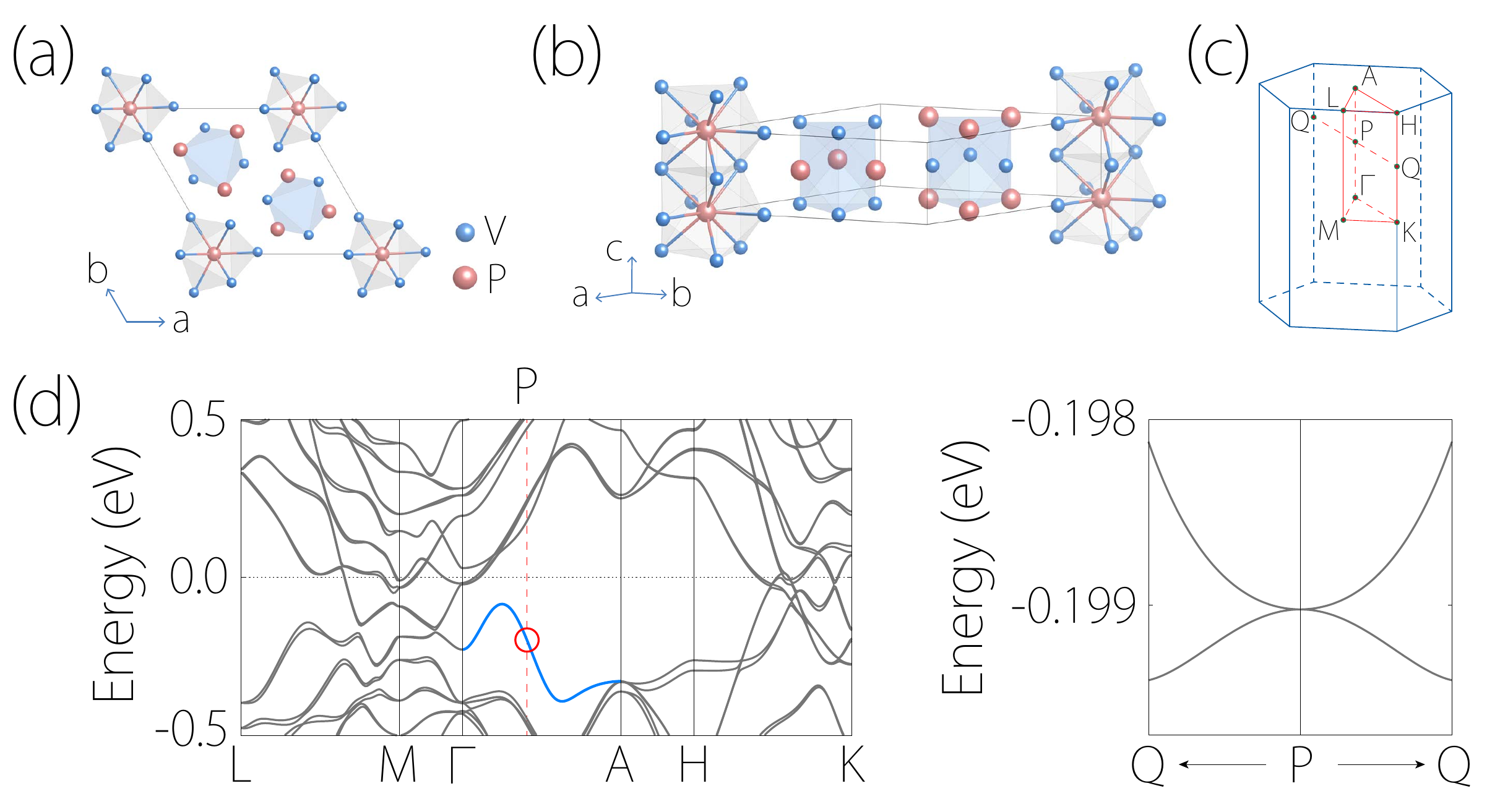}
	\caption{(a) Top and (b) side view of the crystal structure of V$_{12}$P$_{7}$. (c) The corresponding Brillouin zone. The points P and Q are generic points on the paths $\Gamma$-A and K-H, respectively. (d) Calculated band structure of V$_{12}$P$_{7}$ with SOC. The blue color marks the QNL. The right panel shows the enlarged view of the band structure around a generic point P ($k_{z}=0.408\pi$) on the $\Gamma$-A path along the path Q-P-Q which is perpendicular to $\Gamma$-A. }
	\label{V12P7}
\end{figure*}

In the main text, we have shown the result for ZrPtGa, which hosts a QNL along the $\Gamma$-A path. Here, we present a few more examples with the higher-order NLs.
The first example is the QNL material V$_{12}$P$_{7}$, with the structure in the space group $P\bar{6}$ (No.~174), as illustrated in Fig.~\ref{V12P7}(a,b). This material is predicted in Materials Project~\cite{Jain}. The calculated band structure of V$_{12}$P$_{7}$ in the presence of SOC is displayed in Fig.~\ref{V12P7}(d). From the result, one observes a nodal line along the $\Gamma$-A path, about 0.2~eV below the Fermi level. This is a QNL. We purposely show the dispersion along a path P-Q which is perpendicular to the $k_{z}$-axis and P ($k_{z}=0.408\pi$) is a generic point on the $\Gamma$-A path [see the right panel of Fig.~\ref{V12P7}(d)]. The quadratic dispersion is in agreement with our symmetry analysis in the main text. \\


\begin{figure*}
\centering{}\includegraphics[width=8cm]{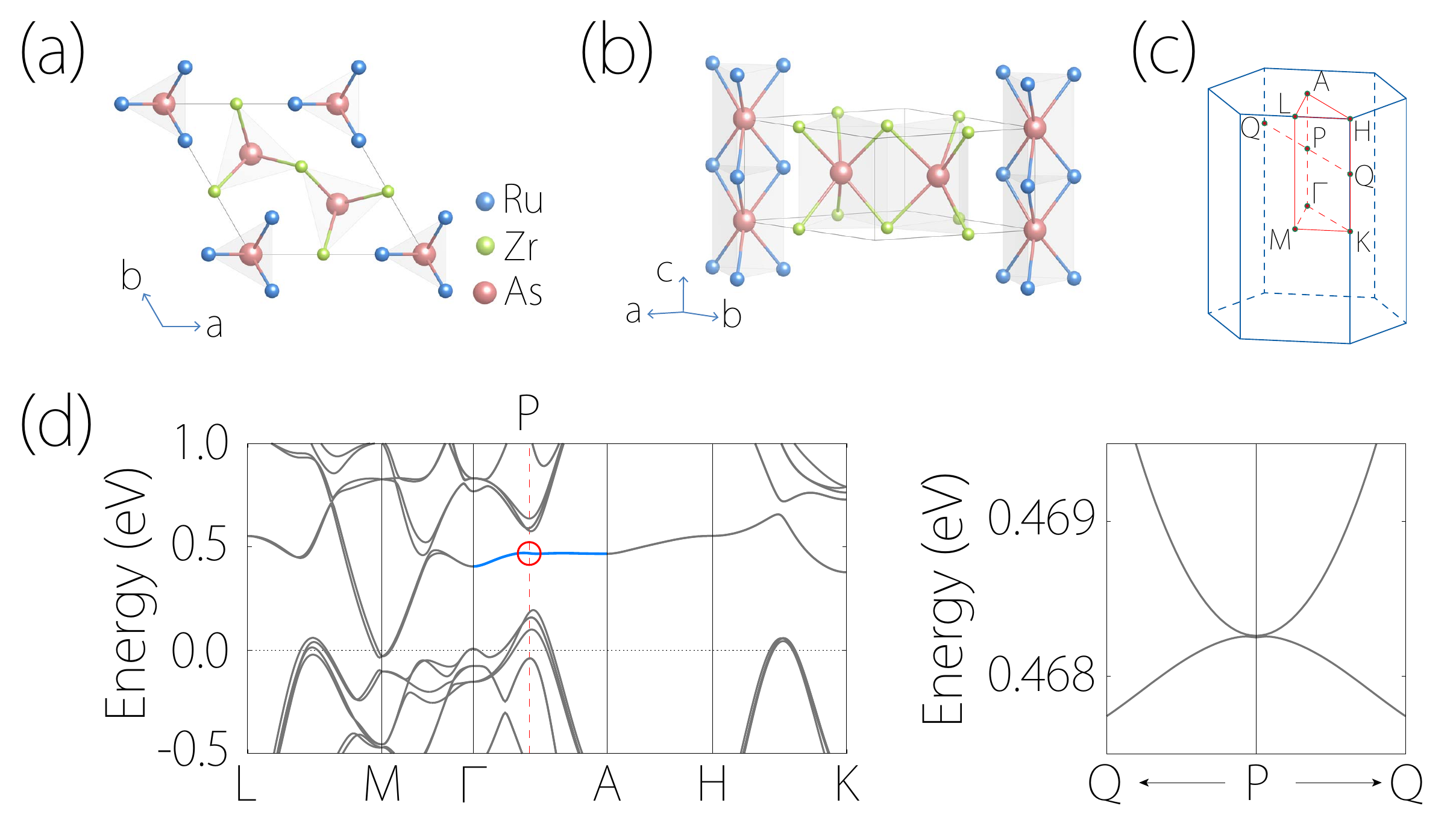}
	\caption{(a) Top and (b) side view of the crystal structure of ZrRuAs. (c) The corresponding Brillouin zone. P and Q are generic points on the paths $\Gamma$-A and K-H, respectively. (d) Calculated band structure of ZrRuAs with SOC. The blue color marks the QNL. The right panel shows the enlarged view of the band structure around a generic point P ($k_{z}=0.41\pi$) on the $\Gamma$-A path along the path Q-P-Q which is perpendicular to $\Gamma$-A path. }
	\label{ZrRuAs}
\end{figure*}

The second example is the QNL material ZrRuAs, which has a crystal structure with the space group of $P\bar{6}2m$ (No.~189) shown in Fig.~\ref{ZrRuAs}(a) and \ref{ZrRuAs}(b). In the band structure for ZrRuAs (with SOC) plotted in Fig.~\ref{ZrRuAs}(d), one observes that a nodal line exists about 0.5~eV above the Fermi level along the $\Gamma$-A path. The dispersion along a generic path P-Q around the nodal line (where P is a generic point $k_{z}=0.41\pi$ on $\Gamma$-A) is shown in the right panel of Fig.~\ref{ZrRuAs}(d). The analysis shows that it is a QNL. \\


\begin{figure*}
\centering{}\includegraphics[width=8cm]{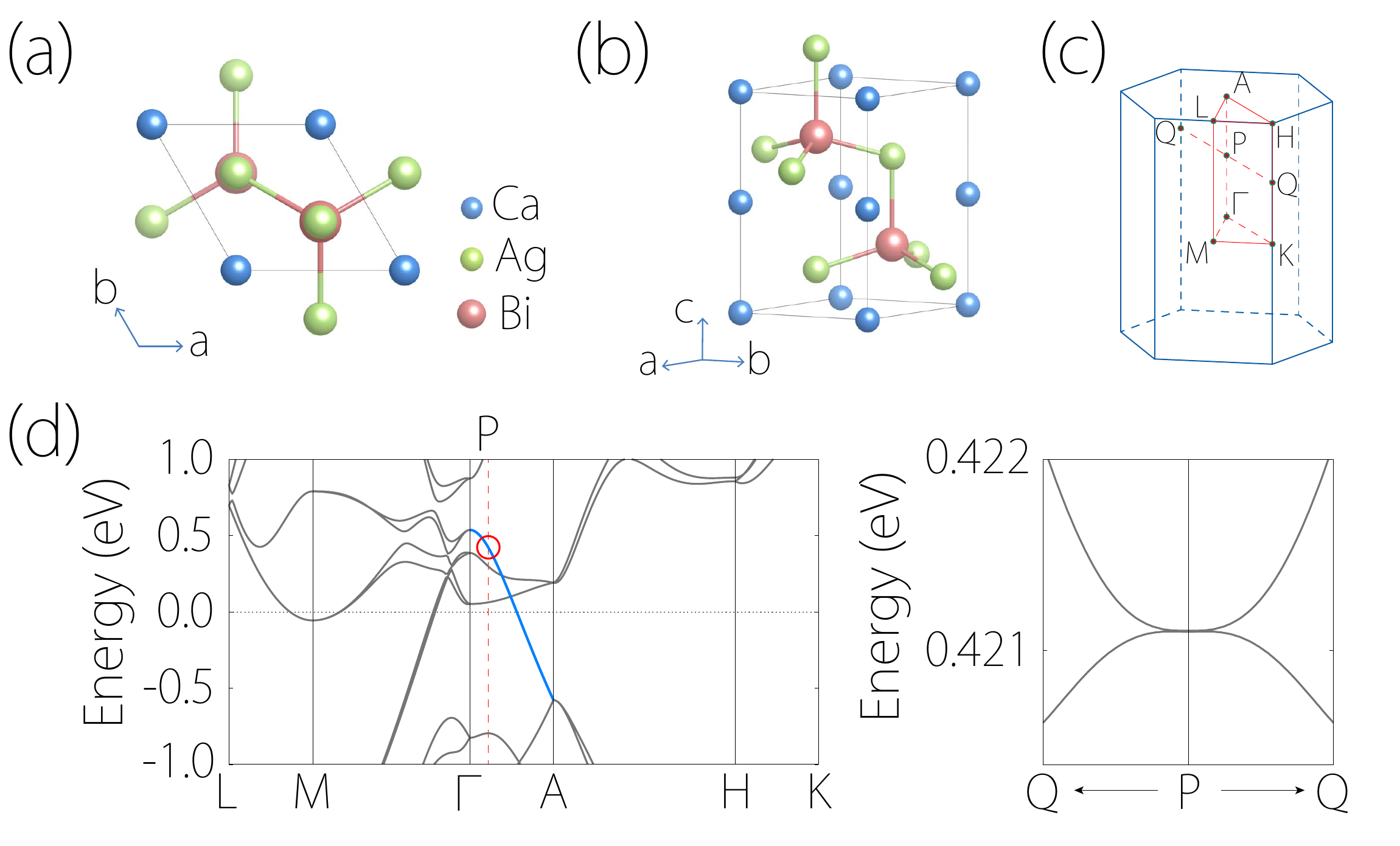}
	\caption{(a) Top and (b) side view of the crystal structure of CaAgBi. (c) The corresponding Brillouin zone. P and Q are generic points on the paths $\Gamma$-A and K-H, respectively. (d) Calculated band structure of CaAgBi with SOC. The cubic nodal line is marked with blue color. The right panel shows the enlarged view of the band structure around a generic point P ($k_{z}=0.222\pi$) on the $\Gamma$-A path along the path Q-P-Q which is perpendicular to $\Gamma$-A path. }
	\label{CaAgBi}
\end{figure*}

The third example is the CNL material CaAgBi, with the space group of $P6_{3}mc$ (No.~186). It has been synthesized in experiment~\cite{Sun}. As shown in Figs.~\ref{CaAgBi}(a) and \ref{CaAgBi}(b), the Ag and Bi atoms form a Wurtzite lattice with encapsulated Ca atoms. The calculated band structure for CaAgBi (with SOC included) is plotted in Fig.~\ref{CaAgBi}(d). Along the $\Gamma$-A path, we finds a CNL (marked with blue in the Figure), with the two-dimensional irreducible representation $\Gamma_{9}$ of the $C_{6v}$ double group. By examining the band dispersion around a generic point (such as $k_{z} = 0.222 \pi$) along the $\Gamma$-A path [see the right panel of Fig.~\ref{CaAgBi}(d)], we find that the dispersion is indeed cubic, consistent with our symmetry analysis.

\section{Derivation of effective Hamiltonian }

Here, we present the derivation of the effective Hamiltonians (shown in the main text) in more detail.

\subsection{SG 174}

As discussed in the main text, for SG 177, the symmetry constraints for a point ${\bm K}$ on $\Gamma$-A are from $C_{3z}$ and ${\cal T}M_{z}$. In the basis corresponding to $\Gamma_{4}$ and $\Gamma_{5}$ representation \cite{Bradley1972,Koster1963}, the symmetry operators are represented by
\begin{equation}
C_{3z}=e^{i\sigma_z\pi/3}, \quad {\cal T}M_{z}=-i\sigma_{x}{\cal K},
\end{equation}
where $\sigma_i$ are the Pauli matrices with $i=x,y,z$, and $\mathcal{K}$ denotes the complex conjugate.
Then, the effective $k\cdot p$ Hamiltonian (up to second order) reads ($\bm q$ is in the plane transverse to the NL)
\begin{equation}
{\cal H}_{{\rm eff}}^{174}(\bm{q}) =  w_{0}+w_{1}q^2+\alpha q_{-}^2\sigma_{+}+\alpha^* q_{+}^2\sigma_{-},
\end{equation}
where $q=\sqrt{q_{x}^{2}+q_{y}^{2}}$, and $w_0$, $w_1$ are real parameters, and $\alpha$ is a complex parameter. All parameters depend on ${\bm K}$. The third order terms, which can violate the emergent chiral symmetry, are given by
\begin{equation}\label{third-174}
\mathcal{H}^{(3)}_{174}=a(q_{+}^3+q_{-}^3)\sigma_z+ib(q_{+}^3-q_{-}^3)\sigma_z,
\end{equation}
where $a$ and $b$ are real parameters that depend on ${\bm K}$.

\subsection{SGs 187-190}

Space groups 187-190 have point group $D_{3h}$. The little group on $\Gamma$-A is $C_{3v}$, for which the generators can be chosen as $C_{3z}$ and a vertical mirror $M_{x(y)}$ for SGs~187-188 (SGs~189-190)~\cite{Bradley1972}. Moreover, there also exists the combined symmetry ${\cal T}M_{z}$. The relevant degenerate basis are those of the $\Gamma_{4}$ representation, which
transform like $\left\{ |\frac{1}{2},-\frac{1}{2}\rangle,\ |\frac{1}{2},\frac{1}{2}\rangle\right\} $ under the symmetry operation~\cite{Bradley1972,Koster1963}.
For SGs~187-188, the symmetry operations are represented by
\begin{equation}
C_{3z}=e^{i\sigma_z\pi/3},\quad M_{x}=-i\sigma_{x},\quad {\cal T}M_{z}=-i\sigma_{x}{\cal K},
\end{equation}
which leads to the following effective model (up to second order)
\begin{equation}
{\cal H}_{{\rm eff}}^{187,188}(\bm{q}) =  w_{0}+w_{1}q^{2}+\alpha q_{-}^2\sigma_{+}+\alpha q_{+}^2\sigma_{-}.\label{H187}
\end{equation}
where $w_{i}$ and $\alpha$ are real parameters that depend on ${\bm K}$. Note that $\alpha$ becomes real in order to preserve the additional symmetry $M_{x}$, compared with the case of SG 174. Still because of the additional $M_{x}$, only the first term of Eq.~\eqref{third-174} remains.

For SGs~189-190, instead of $M_x$, we have $M_{y}=-i\sigma_{y}$, and then the effective model (up to second order) reads
\begin{equation}
{\cal H}_{{\rm eff}}^{189,190}(\bm{q}) =  w_{0}+w_{1}q^{2}+i\alpha q_{-}^2\sigma_{+}-i\alpha q_{+}^2\sigma_{-}.\label{H189}
\end{equation}
where $\alpha$ is real as required by $M_y$. For the third order terms, only the second term of Eq.~\eqref{third-174} can preserve $M_y$.

\subsection{SGs 183-186}

Space groups 183-186 have the same point group $C_{6v}$. The little group on $\Gamma$-A has the generators $C_{6z}$ and a mirror plane $M_{x}$. The relevant basis are those corresponding to the $\Gamma_{9}$ representation of the little group, which transform like $\left\{ |\frac{3}{2},-\frac{3}{2}\rangle,\ |\frac{3}{2},\frac{3}{2}\rangle\right\} $~\cite{Bradley1972,Koster1963}. Hence, the symmetry operations are represented as
\begin{eqnarray}
C_{6z}=i\sigma_{z},\  &  & M_{x}=i\sigma_{x},
\end{eqnarray}
Then the $k\cdot p$ Hamiltonian up to the third order is obtained as
\begin{eqnarray}
{\cal H}_{{\rm eff}}^\text{183-186}(\bm{q}) & = & w_{0}+w_{1}q^2+i(\alpha q_{-}^{3}+\beta q_{+}^{3})\sigma_{+}-i(\alpha q_{+}^3+\beta q_{-}^3)\sigma_{-}\,\label{H183}
\end{eqnarray}
where $w_{i}$, $\alpha$ and $\beta$ are real parameters that depend on ${\bm K}$. Except $w(\bm{q})$, the next order violating the emergent chiral symmetry is the sixth, and there is only one  term given by
\begin{equation}
\mathcal{H}^{(6)}_{183-186}=ic(q_+^6-q_{-}^6)\sigma_{z},
\end{equation}
where $c$ is a real parameter that depends on ${\bm K}$.

\section{Derivation of minimal lattice models }

To further demonstrate the existence of QNLs and CNLs, we explicitly construct minimal lattice (tight-binding) models for the nine SGs listed in Table I of the main text. Here, we follow the method by Wieder and Kane~\cite{wieder2016}. The procedure is to write down the representations for the group generators at $\Gamma$ (including their operations on the wave-vector $\bm{k}$), and then enumerate all the symmetry-allowed hopping terms up to a specified lattice range.

Since QNLs and CNLs predicted here are all formed by two bands, the minimal lattice models would be two-band models for the SGs with only symmorphic symmetry operations, namely for SGs~174, 183, 187 and 189. On the other hand, the minimal lattice models for SGs~184-186, 188, and 190 are four-band models, as the corresponding non-symmorphic symmetries entangle bands in group of four for these space groups~\cite{Watanabe2016}.

\begin{figure*}
\centering{}\includegraphics[width=16cm]{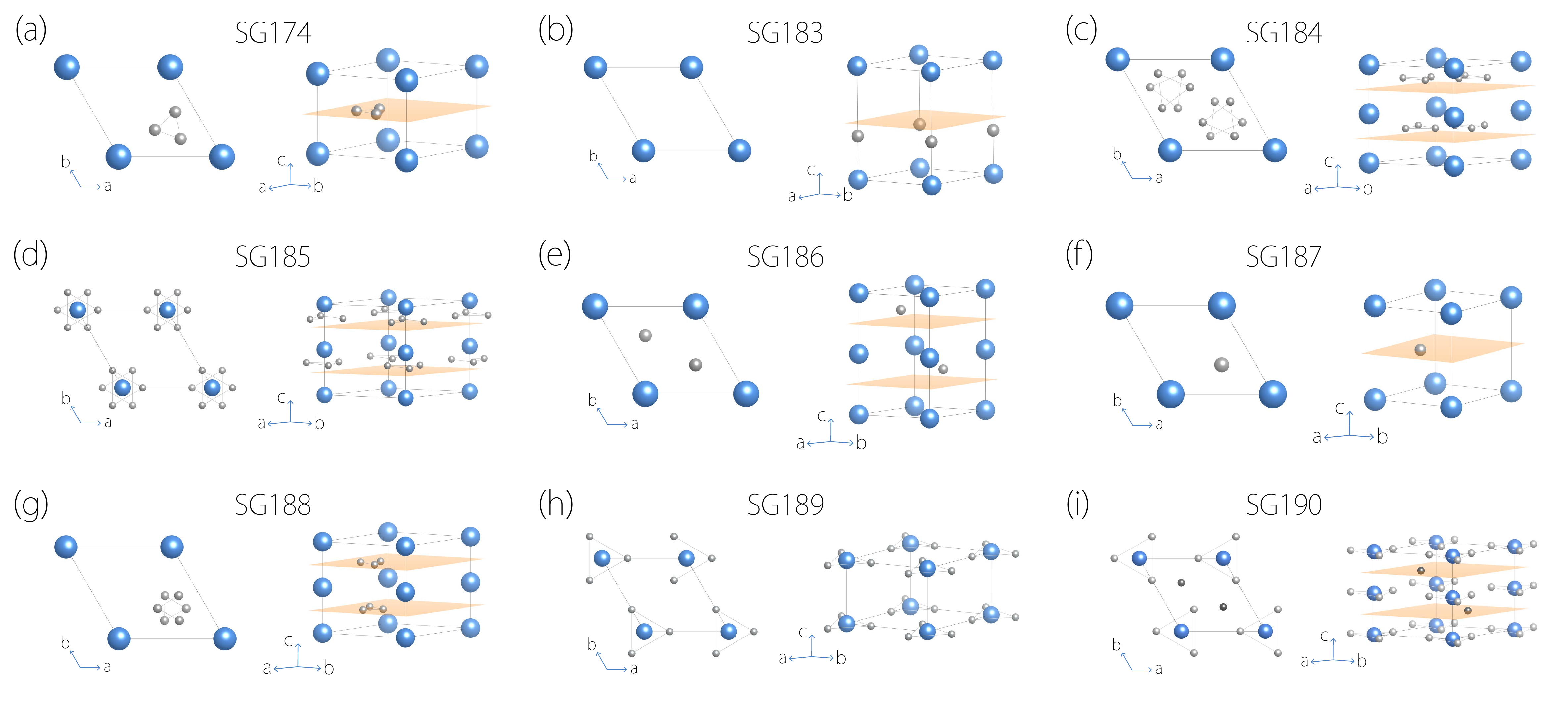}\caption{The structure of real materials belonging to different SGs. Only the atoms (marked by the blue region) sitting on the four equivalent corners
enter in the effective TB model. \label{SI_Fig1}}
\end{figure*}

Specifically, we consider the model defined on an AA-stacked hexagonal lattice, as shown in Fig.~\ref{SI_Fig1}. For (non-)symmorphic SGs, each unit cell contains a single site (two sites) denoted by the blue ball(s), and each site has an $s$-like orbital with two spin states. The gray-colored balls represent sites that do not directly enter the lattice model (because the corresponding orbitals are at high energy), but they affect the hopping amplitudes between the active (blue-colored) sites in a way that follows the SG symmetry. For simplicity, we assume that the lattice
constants $\left(a,b,c\right)$ for symmorphic SGs are the same $c=a=b$, whereas for non-symmorphic SGs are related by $c=2a=2b$.

In the following models, we use two sets of Pauli matrices $\sigma_{i}$ and $\tau_{i}$ to denote operators acting on spin and sublattice spaces, respectively. $\sigma_{0}$  and $\tau_{0}$ both represent the $2\times2$ identity matrix. All the parameters used in the following models are real.

\subsection{SG 174}

The lattice Hamiltonian for SG~174 is constrained by the following symmetries: $C_{3z}$, $M_{z}$, and ${\cal T}$ . With two basis states $\left\{ |\frac{1}{2},-\frac{1}{2}\rangle,\ |\frac{1}{2},\frac{1}{2}\rangle\right\} $ on a single site in a unit cell [see Fig.~\ref{SI_Fig1}(a)], the symmetry operators are represented as
\begin{eqnarray}
C_{3z} & = & e^{i\sigma_z\pi/3}\otimes\left(\begin{array}{cc}
k_{x}\rightarrow\cos\frac{2\pi}{3}k_{x}+\sin\frac{2\pi}{3}k_{y}, & k_{y}\rightarrow-\sin\frac{2\pi}{3}k_{x}+\cos\frac{2\pi}{3}k_{y}\end{array}\right)\\
M_{z} & = & -i\sigma_{z}\otimes\left(k_{z}\rightarrow-k_{z}\right),\\
{\cal T} & = & -i\sigma_{y}{\cal K}\otimes\left(\bm{k}\rightarrow-\bm{k}\right).
\end{eqnarray}
Then the symmetry allowed lattice Hamiltonian can be obtained as
\begin{eqnarray}
H_{\rm TB}^{174} & = & (t_{1}+t_{3}\cos k_{z})\Big(\cos k_{x}+2\cos\frac{k_{x}}{2}\cos\frac{\sqrt{3}k_{y}}{2}\Big)\sigma_{0}+t_{2}\Big(\cos\sqrt{3}k_{y}+2\cos\frac{3k_{x}}{2}\cos\frac{\sqrt{3}k_{y}}{2}\Big)\sigma_{0}+t_{4}\cos k_{z}\sigma_{0}\nonumber \\
 &  & +(t_{1}^{\text{SO}}+t_{3}^{\text{SO}}\cos k_{z})\sin\frac{k_{x}}{2}\Big(\cos\frac{k_{x}}{2}-\cos\frac{\sqrt{3}k_{y}}{2}\Big)\sigma_{z}+t_{2}^{\text{SO}}\sin\frac{\sqrt{3}k_{y}}{2}\Big(\cos\frac{\sqrt{3}k_{y}}{2}-\cos\frac{3k_{x}}{2}\Big)\sigma_{z}\nonumber \\
 &  & +t_{4}^{\text{SO}}\sin k_{z}\Big(\cos k_{x}-\cos\frac{k_{x}}{2}\cos\frac{\sqrt{3}k_{y}}{2}\Big)\sigma_{x}-\sqrt{3}t_{4}^{\text{SO}}\sin k_{z}\sin\frac{k_{x}}{2}\sin\frac{\sqrt{3}k_{y}}{2}\sigma_{y},
\end{eqnarray}
for which we have enumerated all symmetry-allowed terms up to the third-neighbor hopping. The coefficients $t_i$ and $t_i^\text{SO}$ are real valued model parameter. Expanding this model around the $\Gamma$-A path ($k_{x}=k_{y}=0$), we indeed recover the effective Hamiltonian in Eq.~(5) of the main text.

\subsection{SG 183}

The relevant symmetry generators are $C_{6z}$, $M_{x}$, and ${\cal T}$ . With two basis states $\left\{ |\frac{3}{2},-\frac{3}{2}\rangle,\ |\frac{3}{2},\frac{3}{2}\rangle\right\} $ on a single site [see Fig.~\ref{SI_Fig1}(b)], the representations for the symmetry operators are given by
\begin{eqnarray}
C_{6z} & = & i\sigma_{z}\otimes\left(\begin{array}{cc}
k_{x}\rightarrow\cos\frac{\pi}{3}k_{x}+\sin\frac{\pi}{3}k_{y}, & k_{y}\rightarrow-\sin\frac{\pi}{3}k_{x}+\cos\frac{\pi}{3}k_{y}\end{array}\right),\\
M_{x} & = & i\sigma_{x}\otimes\left(k_{x}\rightarrow-k_{x}\right),\\
{\cal T} & = & -i\sigma_{y}{\cal K}\otimes\left(\bm{k}\rightarrow-\bm{k}\right).
\end{eqnarray}
Then the symmetry allowed lattice Hamiltonian can be obtained as
\begin{eqnarray}\label{183}
H_{\rm TB}^{183} & = & \left(t_{1}+t_{3}\cos k_{z}\right)\left(\cos k_{x}+2\cos\frac{k_{x}}{2}\cos\frac{\sqrt{3}k_{y}}{2}\right)\sigma_{0}+\left(t_{2}+t_{4}\cos k_{z}\right)\left(\cos\sqrt{3}k_{y}+2\cos\frac{3k_{x}}{2}\cos\frac{\sqrt{3}k_{y}}{2}\right)\sigma_{0}\nonumber \\
 &  & +t_{5}\cos k_{z}\sigma_{0}+\left(t_{1}^{\text{SO}}+t_{3}^{{\rm SO}}\cos k_{z}\right)\sin\frac{k_{x}}{2}\left(\cos\frac{k_{x}}{2}-\cos\frac{\sqrt{3}}{2}k_{y}\right)\sigma_{y}\nonumber \\
 &  & +\left(t_{2}^{{\rm SO}}+t_{4}^{{\rm SO}}\cos k_{z}\right)\sin\frac{\sqrt{3}k_{y}}{2}\left(\cos\frac{3k_{x}}{2}-\cos\frac{\sqrt{3}}{2}k_{y}\right)\sigma_{x},
 \label{TB183}
\end{eqnarray}
for which we have enumerated all symmetry-allowed terms up to the fourth-neighbor hopping. Expanding this model around the $\Gamma$-A path, we recover the effective Hamiltonian in Eq.~(6) (for the CNL) of the main text.

\subsection{SG 184}

The relevant symmetry generators are $C_{6z}$, the glide mirror plane ${\cal G}_{x}=\left\{ M_{x}|00\frac{1}{2}\right\} $, and ${\cal T}$. Here, we have four basis states per unit cell $\left\{ A|\frac{3}{2},-\frac{3}{2}\rangle,\ A|\frac{3}{2},\frac{3}{2}\rangle,\ B|\frac{3}{2},-\frac{3}{2}\rangle,\ B|\frac{3}{2},\frac{3}{2}\rangle\right\}$, where $A$ and $B$ denote the two sublattices
[see Fig.~\ref{SI_Fig1}(c)]. The symmetry operators are represented as
\begin{eqnarray}
C_{6z} & = & i\sigma_{z}\tau_{0}\otimes\left(\begin{array}{cc}
k_{x}\rightarrow\cos\frac{\pi}{3}k_{x}+\sin\frac{\pi}{3}k_{y}, & k_{y}\rightarrow-\sin\frac{\pi}{3}k_{x}+\cos\frac{\pi}{3}k_{y}\end{array}\right),\\
{\cal G}_{x} & = & i\sigma_{x}\tau_{x}\otimes\left(k_{x}\rightarrow-k_{x}\right),\\
{\cal T} & = & -i\sigma_{y}\tau_{0}{\cal K}\otimes\left(\bm{k}\rightarrow-\bm{k}\right).
\end{eqnarray}
The the symmetry allowed lattice Hamiltonian is obtained as
\begin{eqnarray}
H_{\rm TB}^{184} & = & \left[t_{1}\sigma_{0}\tau_{0}+(t_{3}\sigma_{0}\tau_{x}+t_{5}^{\text{SO}}\sigma_{z}\tau_{y})\cos\frac{k_{z}}{2}\right]\left(\cos k_{x}+2\cos\frac{k_{x}}{2}\cos\frac{\sqrt{3}k_{y}}{2}\right)+\left(t_{4}\sigma_{0}\tau_{x}+t_{6}^{\text{SO}}\sigma_{z}\tau_{y}\right)\cos\frac{k_{z}}{2}\nonumber \\
 &  & +\left[t_{2}\sigma_{0}\tau_{0}+(t_{5}\sigma_{0}\tau_{x}+t_{7}^{\text{SO}}\sigma_{z}\tau_{y})\cos\frac{k_{z}}{2}\right]\left(\cos\sqrt{3}k_{y}+2\cos\frac{3k_{x}}{2}\cos\frac{\sqrt{3}k_{y}}{2}\right)\nonumber \\
 &  & +\left(t_{1}^{\text{SO}}\sigma_{x}\tau_{z}+t_{2}^{\text{SO}}\sigma_{y}\tau_{0}+t_{8}^{\text{SO}}\cos\frac{k_{z}}{2}\sigma_{y}\tau_{x}+t_{9}^{\text{SO}}\sin\frac{k_{z}}{2}\sigma_{x}\tau_{y}\right)\sin\frac{k_{x}}{2}\left(\cos\frac{k_{x}}{2}-\cos\frac{\sqrt{3}k_{y}}{2}\right)\nonumber \\
 &  & +\left(t_{3}^{\text{SO}}\sigma_{x}\tau_{0}+t_{4}^{\text{SO}}\sigma_{y}\tau_{z}+t_{10}^{\text{SO}}\cos\frac{k_{z}}{2}\sigma_{x}\tau_{x}+t_{11}^{\text{SO}}\sin\frac{k_{z}}{2}\sigma_{y}\tau_{y}\right)\sin\frac{\sqrt{3}k_{y}}{2}\left(\cos\frac{3k_{x}}{2}-\cos\frac{\sqrt{3}k_{y}}{2}\right),
\end{eqnarray}
for which we have enumerated all symmetry-allowed terms up to the fourth-neighbor hopping. One can check that the four bands on the $\Gamma$-A path form two CNLs.

\subsection{SG 185}

The relevant symmetry generators are the screw rotation ${\cal S}_{6z}=\left\{ C_{6z}|00\frac{1}{2}\right\} $, $M_{x}$, and ${\cal T}$. Like the previous case, with the basis $\left\{ A|\frac{3}{2},-\frac{3}{2}\rangle,\ A|\frac{3}{2},\frac{3}{2}\rangle,\ B|\frac{3}{2},-\frac{3}{2}\rangle,\ B|\frac{3}{2},\frac{3}{2}\rangle\right\} $ [see Fig.~\ref{SI_Fig1}(d)], the symmetry operators are represented as
\begin{eqnarray}
\mathcal{S}_{6z} & = & i\sigma_{z}\tau_{x}\otimes\left(\begin{array}{cc}
k_{x}\rightarrow\cos\frac{\pi}{3}k_{x}+\sin\frac{\pi}{3}k_{y}, & k_{y}\rightarrow-\sin\frac{\pi}{3}k_{x}+\cos\frac{\pi}{3}k_{y}\end{array}\right),\\
M_{x} & = & i\sigma_{x}\tau_{0}\otimes\left(k_{x}\rightarrow-k_{x}\right),\\
{\cal T} & = & -i\sigma_{y}\tau_{0}{\cal K}\otimes\left(\bm{k}\rightarrow-\bm{k}\right).
\end{eqnarray}
Then the symmetry allowed lattice Hamiltonian can be obtained as
\begin{eqnarray}
H_{\rm TB}^{185} & = & \left[t_{1}\sigma_{0}\tau_{0}+(t_{3}\sigma_{0}\tau_{x}+t_{4}^{\text{SO}}\sigma_{x}\tau_{y})\cos\frac{k_{z}}{2}\right]\left(\cos k_{x}+2\cos\frac{k_{x}}{2}\cos\frac{\sqrt{3}k_{y}}{2}\right)+\left(t_{4}\sigma_{0}\tau_{x}+t_{5}^{\text{SO}}\sigma_{x}\tau_{y}\right)\cos\frac{k_{z}}{2}\nonumber \\
 &  & +\left[t_{2}\sigma_{0}\tau_{0}+(t_{5}\sigma_{0}\tau_{x}+t_{6}^{\text{SO}}\sigma_{x}\tau_{y})\cos\frac{k_{z}}{2}\right]\left(\cos\sqrt{3}k_{y}+2\cos\frac{3k_{x}}{2}\cos\frac{\sqrt{3}k_{y}}{2}\right)\nonumber \\
 &  & +\left[t_{1}^{\text{SO}}\sigma_{x}\tau_{0}+\left(t_{6}\sigma_{0}\tau_{y}+t_{7}^{\text{SO}}\sigma_{x}\tau_{x}\right)\cos\frac{k_{z}}{2}\right]\sin\frac{\sqrt{3}k_{y}}{2}\left(\cos\frac{3k_{x}}{2}-\cos\frac{\sqrt{3}k_{y}}{2}\right)\nonumber \\
 &  & +\left(t_{2}^{\text{SO}}\sigma_{z}\tau_{z}+t_{3}^{\text{SO}}\sigma_{y}\tau_{0}+t_{8}^{\text{SO}}\sin\frac{k_{z}}{2}\sigma_{z}\tau_{y}+t_{9}^{\text{SO}}\cos\frac{k_{z}}{2}\sigma_{y}\tau_{x}\right)\sin\frac{k_{x}}{2}\left(\cos\frac{k_{x}}{2}-\cos\frac{\sqrt{3}k_{y}}{2}\right),
\end{eqnarray}
for which we have enumerated all symmetry-allowed terms up to the fourth-neighbor hopping. The Hamiltonian $H_{\rm TB}^{185}$ contains two CNLs on the $\Gamma$-A path. 

\subsection{SG 186}

The relevant symmetry generators are the screw rotation
${\cal S}_{6z}=\left\{ C_{6z}|00\frac{1}{2}\right\} $,  the glide mirror ${\cal G}_{x}=\left\{ M_{x}|00\frac{1}{2}\right\} $,
and ${\cal T}$. With the basis $\left\{ A|\frac{3}{2},-\frac{3}{2}\rangle,\ A|\frac{3}{2},\frac{3}{2}\rangle,\ B|\frac{3}{2},-\frac{3}{2}\rangle,\ B|\frac{3}{2},\frac{3}{2}\rangle\right\} $ [see Fig.~\ref{SI_Fig1}(e)], the representations for the symmetry operators are
\begin{eqnarray}
\mathcal{S}_{6z} & = & i\sigma_{z}\tau_{x}\otimes\left(\begin{array}{cc}
k_{x}\rightarrow\cos\frac{\pi}{3}k_{x}+\sin\frac{\pi}{3}k_{y}, & k_{y}\rightarrow-\sin\frac{\pi}{3}k_{x}+\cos\frac{\pi}{3}k_{y}\end{array}\right),\\
\mathcal{G}_{x} & = & i\sigma_{x}\tau_{x}\otimes\left(k_{x}\rightarrow-k_{x}\right),\\
{\cal T} & = & -i\sigma_{y}\tau_{0}{\cal K}\otimes\left(\bm{k}\rightarrow-\bm{k}\right).
\end{eqnarray}
Then the symmetry allowed lattice Hamiltonian can be obtained as
\begin{eqnarray}
H_{\rm TB}^{186} & = & \left[t_{1}\sigma_{0}\tau_{0}+(t_{3}\sigma_{0}\tau_{x}+t_{4}^{\text{SO}}\sigma_{y}\tau_{y})\cos\frac{k_{z}}{2}\right]\left(\cos k_{x}+2\cos\frac{k_{x}}{2}\cos\frac{\sqrt{3}k_{y}}{2}\right)+\left(t_{4}\sigma_{0}\tau_{x}+t_{5}^{\text{SO}}\sigma_{y}\tau_{y}\right)\cos\frac{k_{z}}{2}\nonumber \\
 &  & +\left[t_{2}\sigma_{0}\tau_{0}+(t_{5}\sigma_{0}\tau_{x}+t_{6}^{\text{SO}}\sigma_{y}\tau_{y})\cos\frac{k_{z}}{2}\right]\left(\cos\sqrt{3}k_{y}+2\cos\frac{3k_{x}}{2}\cos\frac{\sqrt{3}k_{y}}{2}\right)\nonumber \\
 &  & +\left[t_{1}^{\text{SO}}\sigma_{y}\tau_{0}+(t_{6}\sigma_{0}\tau_{y}+t_{6}^{\text{SO}}\sigma_{y}\tau_{x})\cos\frac{k_{z}}{2}\right]\sin\frac{k_{x}}{2}\left(\cos\frac{k_{x}}{2}-\cos\frac{\sqrt{3}k_{y}}{2}\right)\nonumber \\
 &  & +\left[t_{2}^{\text{SO}}\sigma_{z}\tau_{z}+t_{3}^{\text{SO}}\sigma_{x}\tau_{0}+t_{8}^{\text{SO}}\cos\frac{k_{z}}{2}\sigma_{x}\tau_{x}+t_{9}^{\text{SO}}\sin\frac{k_{z}}{2}\sigma_{z}\tau_{y}\right]\sin\frac{\sqrt{3}k_{y}}{2}\left(\cos\frac{\sqrt{3}k_{y}}{2}-\cos\frac{3k_{x}}{2}\right),
\end{eqnarray}
for which we have enumerated all symmetry-allowed terms up to the fourth-neighbor hopping. The Hamiltonian $H_{\rm TB}^{186}$ has two CNLs on the $\Gamma$-A path. 

\subsection{SG 187}

The relevant symmetry generators are $C_{3z}$, $M_{x}$, $M_{z}$, and ${\cal T}$. With the basis $\left\{ |\frac{1}{2},-\frac{1}{2}\rangle,\ |\frac{1}{2},\frac{1}{2}\rangle\right\} $ on a single site in the unit cell [see Fig.~\ref{SI_Fig1}(f)], the representations for the symmetry operators are
\begin{eqnarray}
C_{3z} & = & e^{i\sigma_z\pi/3}\otimes\left(\begin{array}{cc}
k_{x}\rightarrow\cos\frac{2\pi}{3}k_{x}+\sin\frac{2\pi}{3}k_{y}, & k_{y}\rightarrow-\sin\frac{2\pi}{3}k_{x}+\cos\frac{2\pi}{3}k_{y}\end{array}\right),\\
M_{x} & = & -i\sigma_{x}\otimes\left(k_{x}\rightarrow-k_{x}\right),\\
M_{z} & = & -i\sigma_{z}\otimes\left(k_{z}\rightarrow-k_{z}\right),\\
{\cal T} & = & -i\sigma_{y}{\cal K}\otimes\left(\bm{k}\rightarrow-\bm{k}\right).
\end{eqnarray}
Then the symmetry allowed lattice Hamiltonian can be obtained as
\begin{eqnarray}\label{187}
H_{\rm TB}^{187} & = & (t_{1}+t_{2}\cos k_{z})\left(\cos k_{x}+2\cos\frac{k_{x}}{2}\cos\frac{\sqrt{3}k_{y}}{2}\right)\sigma_{0}+(t_{1}^{\text{SO}}+t_{2}^{\text{SO}}\cos k_{z})\sin\frac{k_{x}}{2}\left(\cos\frac{k_{x}}{2}-\cos\frac{\sqrt{3}k_{y}}{2}\right)\sigma_{z}\nonumber \\
 &  & +t_{3}\cos k_{z}\sigma_{0}+t_{3}^{\text{SO}}\sin k_{z}\left[\left(\cos k_{x}-\cos\frac{k_{x}}{2}\cos\frac{\sqrt{3}k_{y}}{2}\right)\sigma_{x}-\sqrt{3}\sin\frac{k_{x}}{2}\sin\frac{\sqrt{3}k_{y}}{2}\sigma_{y}\right],\label{TB187}
\end{eqnarray}
for which we have enumerated all symmetry-allowed terms up to the second-neighbor hopping. Expanding this model around the $\Gamma$-A path, we recover the effective Hamiltonian for QNLs in the main text.

\subsection{SG 188}

The relevant symmetry generators are $C_{3z}$, ${\cal G}_{x}=\{M_{x}|00\frac{1}{2}\}$, $M_{z}$, and ${\cal T}$. With the four-states basis $\left\{ A|\frac{1}{2},-\frac{1}{2}\rangle,\ A|\frac{1}{2},\frac{1}{2}\rangle,\ B|\frac{1}{2},-\frac{1}{2}\rangle,\ B|\frac{1}{2},\frac{1}{2}\rangle\right\}$
[see Fig.~\ref{SI_Fig1}(g)], the representations for the symmetry operators are
\begin{eqnarray}
C_{3z} & = & e^{i\sigma_z\pi/3}\tau_{0}\otimes\left(\begin{array}{cc}
k_{x}\rightarrow\cos\frac{2\pi}{3}k_{x}+\sin\frac{2\pi}{3}k_{y}, & k_{y}\rightarrow-\sin\frac{2\pi}{3}k_{x}+\cos\frac{2\pi}{3}k_{y}\end{array}\right),\\
{\cal G}_{x}  & = & -i\sigma_{x}\tau_{x}\otimes\left(k_{x}\rightarrow-k_{x}\right),\\
M_{z} & = & -i\sigma_{z}\tau_{0}\otimes\left(k_{z}\rightarrow-k_{z}\right),\\
{\cal T} & = & -i\sigma_{y}\tau_{0}{\cal K}\otimes\left(\bm{k}\rightarrow-\bm{k}\right).
\end{eqnarray}

Then the symmetry allowed lattice Hamiltonian can be obtained as
\begin{eqnarray}
H_{\rm TB}^{188} & = & \left[t_{1}\sigma_{0}\tau_{0}+(t_{2}\sigma_{0}\tau_{x}+t_{2}^{\text{SO}}\sigma_{z}\tau_{y})\cos\frac{k_{z}}{2}\right]\left(\cos k_{x}+2\cos\frac{k_{x}}{2}\cos\frac{\sqrt{3}k_{y}}{2}\right)+\left(t_{3}\sigma_{0}\tau_{x}+t_{3}^{\text{SO}}\sigma_{z}\tau_{y}\right)\cos\frac{k_{z}}{2}\nonumber \\
 &  & +\left[t_{1}^{\text{SO}}\sigma_{z}\tau_{0}+(t_{4}\sigma_{0}\tau_{y}+t_{4}^{\text{SO}}\sigma_{z}\tau_{x})\cos\frac{k_{z}}{2}\right]\sin\frac{k_{x}}{2}\left(\cos\frac{k_{x}}{2}-\cos\frac{\sqrt{3}k_{y}}{2}\right)\nonumber \\
 &  & +t_{5}^{\text{SO}}\sin\frac{k_{z}}{2}\left[\left(\cos k_{x}-\cos\frac{k_{x}}{2}\cos\frac{\sqrt{3}k_{y}}{2}\right)\sigma_{x}\tau_{x}-\sqrt{3}\sin\frac{k_{x}}{2}\sin\frac{\sqrt{3}k_{y}}{2}\sigma_{y}\tau_{x}\right]\nonumber \\
 &  & +t_{6}^{\text{SO}}\sin\frac{k_{z}}{2}\left[\left(\sin k_{x}+\sin\frac{k_{x}}{2}\cos\frac{\sqrt{3}k_{y}}{2}\right)\sigma_{x}\tau_{y}-\sqrt{3}\cos\frac{k_{x}}{2}\sin\frac{\sqrt{3}k_{y}}{2}\sigma_{y}\tau_{y}\right],
\end{eqnarray}
for which we have enumerated all symmetry-allowed terms up to the second-neighbor hopping. Expanding this model around the $\Gamma$-A path, we recover the effective Hamiltonian for QNLs in the main text.

\subsection{SG 189}

The relevant symmetry generators are $C_{3z}$, $M_{y}$, $M_{z}$, and ${\cal T}$. With the basis $\left\{ |\frac{1}{2},-\frac{1}{2}\rangle,\ |\frac{1}{2},\frac{1}{2}\rangle\right\} $ [see Fig.~\ref{SI_Fig1}(h)], the representations for the symmetry operators are
\begin{eqnarray}
C_{3z} & = & e^{i\sigma_z\pi/3}\otimes\left(\begin{array}{cc}
k_{x}\rightarrow\cos\frac{2\pi}{3}k_{x}+\sin\frac{2\pi}{3}k_{y}, & k_{y}\rightarrow-\sin\frac{2\pi}{3}k_{x}+\cos\frac{2\pi}{3}k_{y}\end{array}\right),\\
M_{y} & = & -i\sigma_{y}\otimes\left(k_{y}\rightarrow-k_{y}\right),\\
M_{z} & = & -i\sigma_{z}\otimes\left(k_{z}\rightarrow-k_{z}\right),\\
{\cal T} & = & -i\sigma_{y}{\cal K}\otimes\left(\bm{k}\rightarrow-\bm{k}\right).
\end{eqnarray}
Then the symmetry allowed lattice Hamiltonian can be obtained as
\begin{eqnarray}
H_{\rm TB}^{189} & = & (t_{1}+t_{3}\cos k_{z})\left(\cos k_{x}+2\cos\frac{k_{x}}{2}\cos\frac{\sqrt{3}k_{y}}{2}\right)\sigma_{0}+(t_{2}+t_{4}\cos k_{z})\left(\cos\sqrt{3}k_{y}+2\cos\frac{3k_{x}}{2}\cos\frac{\sqrt{3}k_{y}}{2}\right)\sigma_{0}\nonumber \\
 &  & +t_{5}\cos k_{z}\sigma_{0}+(t_{1}^{\text{SO}}+t_{2}^{\text{SO}}\cos k_{z})\sin\frac{\sqrt{3}k_{y}}{2}\left(\cos\frac{\sqrt{3}k_{y}}{2}-\cos\frac{3k_{x}}{2}\right)\sigma_{z}\nonumber \\
 &  & +t_{3}^{\text{SO}}\sin k_{z}\left[\sqrt{3}\sin\frac{k_{x}}{2}\sin\frac{\sqrt{3}k_{y}}{2}\sigma_{x}+\left(\cos k_{x}-\cos\frac{k_{x}}{2}\cos\frac{\sqrt{3}k_{y}}{2}\right)\sigma_{y}\right]\nonumber \\
 &  & +t_{4}^{\text{SO}}\sin k_{z}\left[\sqrt{3}\sin\frac{3k_{x}}{2}\sin\frac{\sqrt{3}k_{y}}{2}\sigma_{x}-\left(\cos\sqrt{3}k_{y}-\cos\frac{3k_{x}}{2}\cos\frac{\sqrt{3}k_{y}}{2}\right)\sigma_{y}\right],
\end{eqnarray}
for which we have enumerated all symmetry-allowed terms up to the fourth-neighbor hopping. Expanding this model around the $\Gamma$-A path, we recover the effective Hamiltonian for QNLs in the main text.

\subsection{SG 190}

The relevant symmetry generators are $C_{3z}$, ${\cal G}_{y}=\{M_{y}|00\frac{1}{2}\}$,
$M_{z}$, and ${\cal T}$. With the four-state basis $\left\{ A|\frac{1}{2},-\frac{1}{2}\rangle,\ A|\frac{1}{2},\frac{1}{2}\rangle,\ B|\frac{1}{2},-\frac{1}{2}\rangle,\ B|\frac{1}{2},\frac{1}{2}\rangle\right\}$
[see Fig.~\ref{SI_Fig1}(i)], the representations for the symmetry operators are
\begin{eqnarray}
C_{3z} & = & e^{i\sigma_z\pi/3}\tau_{0}\otimes\left(\begin{array}{cc}
k_{x}\rightarrow\cos\frac{2\pi}{3}k_{x}+\sin\frac{2\pi}{3}k_{y}, & k_{y}\rightarrow-\sin\frac{2\pi}{3}k_{x}+\cos\frac{2\pi}{3}k_{y}\end{array}\right),\\
{\cal G}_{y}  & = & -i\sigma_{y}\tau_{x}\otimes\left(k_{y}\rightarrow-k_{y}\right),\\
M_{z} & = & -i\sigma_{z}\tau_{0}\otimes\left(k_{z}\rightarrow-k_{z}\right),\\
{\cal T} & = & -i\sigma_{y}\tau_{0}{\cal K}\otimes\left(\bm{k}\rightarrow-\bm{k}\right).
\end{eqnarray}
Then the symmetry allowed lattice Hamiltonian can be obtained as
\begin{eqnarray}
H_{\rm TB}^{190} & = & \left[t_{1}\sigma_{0}\tau_{0}+\left(t_{3}\sigma_{0}\tau_{x}+t_{3}^{\text{SO}}\sigma_{z}\tau_{y}\right)\cos\frac{k_{z}}{2}\right]\left(\cos k_{x}+2\cos\frac{k_{x}}{2}\cos\frac{\sqrt{3}k_{y}}{2}\right)+t_{1}^{\text{SO}}\sin\frac{k_{x}}{2}\left(\cos\frac{k_{x}}{2}-\cos\frac{\sqrt{3}k_{y}}{2}\right)\sigma_{z}\tau_{z}\nonumber \\
 &  & +\left[t_{2}\sigma_{0}\tau_{0}+\left(t_{5}\sigma_{0}\tau_{x}+t_{5}^{\text{SO}}\sigma_{z}\tau_{y}\right)\cos\frac{k_{z}}{2}\right]\left(\cos\sqrt{3}k_{y}+2\cos\frac{3k_{x}}{2}\cos\frac{\sqrt{3}k_{y}}{2}\right)+\left(t_{4}\sigma_{0}\tau_{x}+t_{4}^{\text{SO}}\sigma_{z}\tau_{y}\right)\cos\frac{k_{z}}{2}\nonumber \\
 &  & +\left[t_{2}^{\text{SO}}\sigma_{z}\tau_{0}+\left(t_{6}\sigma_{0}\tau_{y}+t_{6}^{\text{SO}}\sigma_{z}\tau_{x}\right)\cos\frac{k_{z}}{2}\right]\sin\frac{\sqrt{3}k_{y}}{2}\left(\cos\frac{\sqrt{3}k_{y}}{2}-\cos\frac{3k_{x}}{2}\right)\nonumber \\
 &  & +t_{7}^{\text{SO}}\sin\frac{k_{z}}{2}\left[\sqrt{3}\cos\frac{k_{x}}{2}\sin\frac{\sqrt{3}k_{y}}{2}\sigma_{y}\tau_{y}-\sin\frac{k_{x}}{2}\left(2\cos\frac{k_{x}}{2}+\cos\frac{\sqrt{3}k_{y}}{2}\right)\sigma_{x}\tau_{y}\right]\nonumber \\
 &  & +t_{8}^{\text{SO}}\sin\frac{k_{z}}{2}\left[\sqrt{3}\sin\frac{k_{x}}{2}\sin\frac{\sqrt{3}k_{y}}{2}\sigma_{x}\tau_{x}+\left(\cos k_{x}-\cos\frac{k_{x}}{2}\cos\frac{\sqrt{3}k_{y}}{2}\right)\sigma_{y}\tau_{x}\right]\nonumber \\
 &  & +t_{9}^{\text{SO}}\sin\frac{k_{z}}{2}\left[\sqrt{3}\sin\frac{3k_{x}}{2}\sin\frac{\sqrt{3}k_{y}}{2}\sigma_{x}\tau_{x}-\left(\cos\sqrt{3}k_{y}-\cos\frac{3k_{x}}{2}\cos\frac{\sqrt{3}k_{y}}{2}\right)\sigma_{y}\tau_{x}\right]\nonumber \\
 &  & +t_{10}^{\text{SO}}\sin\frac{k_{z}}{2}\left[\sqrt{3}\sin\frac{3k_{x}}{2}\cos\frac{\sqrt{3}k_{y}}{2}\sigma_{x}\tau_{y}-\sin\frac{\sqrt{3}k_{y}}{2}\left(\cos\frac{3k_{x}}{2}+2\cos\frac{\sqrt{3}k_{y}}{2}\right)\sigma_{y}\tau_{y}\right],
\end{eqnarray}
for which we have enumerated all symmetry-allowed terms up to the fourth-neighbor hopping. Expanding this model around the $\Gamma$-A path, we recover the effective Hamiltonian for QNLs in the main text.

\section{Zero Landau levels for higher-order NL}

The topological charge $\mathcal{N}$ defined in the main text also determines the number of zero Landau levels (LLs) for the effective model $\mathcal{H}_\text{eff}$. This result has been proved in Ref.~\cite{Landau-Lattice}. Here, we have already derived the result explicitly for $\mathcal{H}_\text{eff}$ for QNLs and CNLs in Table~I of the main text. To get the Hamiltonian with magnetic field, we simply replace $q_\pm$ by $a^\dagger$ and $a$, respectively, where $a^\dagger$ and $a$ are boson operators corresponding to the cyclotron-motion under a magnetic field. For instance, $\mathcal{H}_\text{eff}$ for the QNL in SG~174 has topological charge $\mathcal{N}=2$, hence the number of zero LLs is accordingly two. Indeed, the Hamiltonian is $\mathcal{H}^{174}_\text{LL}=\alpha a^2 \sigma_++\mathrm{H.c.}$, for which $|0,\downarrow\rangle$ and $|1,\downarrow\rangle$ are two degenerate states with zero energy. It is also worth noting that higher-order terms that break the emergent chiral symmetry for $\mathcal{H}_\text{eff}$ will not remove the zero LLs but will generally split the degeneracy and shift the LLs from zero energy.

\section{Calculation details for Fig.~2 of main text}

In Fig.~2 of the main text, we studied the surface states of a lattice model corresponding to SG~183 which contains a CNL. The model we take is Eq.~(\ref{183}) which we derive for SG~183. In calculating the bulk and surface band structures in Fig.~2, we take the following model parameters: $t_1=0.1$, $t_2=0.01$, $t_3=-0.04$, $t_4=0.004$, $t_5=0.2$, $t_{1}^{\text{SO}}=-4$, $t_{2}^{\text{SO}}=-3.85$, $t_{3}^{\text{SO}}=-0.8$ and $t_{4}^{\text{SO}}=0.77$. The calculation in Fig.~2(c) is performed for a slab with a thickness of 35 layers of lattice sites.

\section{Calculation details for Fig.~3 of main text}

Here we present the calculation details for Fig.~3 of the main text. In Fig.~3(a), we consider breaking $C_{3z}$ for a system that contains a QNL. Here, to be specific, we take the lattice model Eq.~(\ref{187}) for SG~187, and break the $C_{3z}$ symmetry by adding a perturbation term
\begin{eqnarray}
\Delta H & = & \delta t_{1}^{{\rm SO}} \sin k_z\ \sigma_x.
\end{eqnarray}
Then we find that the QNL is split into two linear NLs schematically shown in Fig.~3(a).

In Fig.~3(b), we again take the model Eq.~(\ref{187}) for SG~187, and break the $M_z$ symmetry by adding the the following perturbation term
\begin{eqnarray}
\Delta H & = &\delta t_{2}^{{\rm SO}}\left[3\cos\frac{k_{x}}{2}\sin\frac{\sqrt{3}k_{y}}{2}\cos k_{z}+\left(\cos\frac{k_{x}}{2}\cos\frac{\sqrt{3}k_{y}}{2}-\cos k_{x}\right)\sin k_{z}\right]\sigma_{x} \nonumber\\
 &  & +\sqrt{3}\delta t_{2}^{{\rm SO}}\left[\sin\frac{k_{x}}{2}\sin\frac{\sqrt{3}k_{y}}{2}\sin k_{z}+\left(\sin\frac{k_{x}}{2}\cos\frac{\sqrt{3}k_{y}}{2}+\sin k_{x}\right)\cos k_{z}\right]\sigma_{y}.
\end{eqnarray}
Then we find that the QNL is transformed to four intertwined linear NLs as schematically illustrated in Fig.~3(b).

In Fig.~3(c), we consider breaking the vertical mirror for a system with CNL. First, we show that generally the CNL will be gapped and there will appear a triple Weyl point at $\Gamma$. This can be done by constructing a $k\cdot p$ Hamiltonian around the
$\Gamma$ point with the remaining symmetry constraints of $C_{3z}$ and ${\cal T}$. In the basis of $\left\{ |\frac{3}{2},-\frac{3}{2}\rangle,\ |\frac{3}{2},\frac{3}{2}\rangle\right\} $, we find that up to the third order, the effective Hamiltonian reads
\begin{eqnarray}
{\cal H}_{{\rm eff}} & = & \sum_{m=x,y,z}\left[d_{m,1}k_{z}(1+d_{m,2}k_{\parallel}^{2}+d_{m,3}k_{z}^{2})+d_{m,4}K_{+}+d_{m,5}K_{-}\right]\sigma_{m},
\end{eqnarray}
where $k_\|=\sqrt{k_x^2+k_y^2}$, $K_{+}=(k_{x}+ik_{y})^{3}+(k_{x}-ik_{y})^{3}$ and $K_{-}=i\left[(k_{x}+ik_{y})^{3}-(k_{x}-ik_{y})^{3}\right]$.
One can check that such Weyl point possesses a cubic dispersion in the $k_{x}$-$k_{y}$ plane and a linear dispersion in the $k_{z}$ direction, and it carries a Chern number of $\pm 3$. This is verified by the lattice model calculation. We take the model (\ref{TB183}) for SG~183, and add a perturbation term
\begin{eqnarray}
\Delta H & = &\delta t_{1}^{{\rm SO}}\sin k_z\ \sigma_{z},
\end{eqnarray}
which breaks the $M_x$ symmetry. Then we indeed find that there are two triple Weyl points located at $\Gamma$ and at $A$, as illustrated in Fig.~3(c) of the main text.